\documentclass[a4paper, 11pt]{article}
\usepackage{jheppub}
\usepackage[T1]{fontenc} 
\usepackage{graphicx}
\usepackage{tikz}
\usepackage{amsmath,amssymb,multirow,xspace,slashed,array,booktabs}
\usepackage[colorlinks=true,urlcolor=blue,anchorcolor=blue,citecolor=blue,filecolor=blue,linkcolor=blue,menucolor=blue,pagecolor=blue]{hyperref}
\usepackage{natbib}
\setcitestyle{square, comma, numbers,sort&compress, super}
\usepackage{booktabs}
\usepackage{blindtext, rotating}
\usepackage{appendix}
\usepackage{afterpage}
\usepackage{orcidlink}
\usepackage{changepage}
\usepackage{enumitem}
\usepackage{marvosym}
\usepackage{bbold}
\usepackage{slashed}
\usepackage{physics}
\usepackage{verbatim}
\usepackage{indentfirst}
\usepackage{bm}
\usepackage{soul} 
\usepackage[normalem]{ulem}
\usepackage{pifont}
\usepackage{cancel}
\usepackage{float}
\usepackage{caption}
\usepackage{subcaption}
\usepackage{cleveref}
\crefname{figure}{Fig.}{Figs.}
\Crefname{figure}{Fig.}{Figs.}

\crefrangeformat{figure}{Figs.~#3#1#4--#5#2#6}
\Crefrangeformat{figure}{Figs.~#3#1#4--#5#2#6}

\newcommand{\gev}{\mathrm{GeV}}

\title{
    Dark Matter and Baryon Asymmetry from Monopole-Axion Interactions
}

\author[a]{Raymond T. Co\,\orcidlink{0000-0002-8395-7056},}
\author[b,c]{Keisuke Harigaya\,\orcidlink{0000-0001-6516-3386},}
\author[d]{Isaac R. Wang\,\orcidlink{0000-0003-0789-218X},}
\author[d,e,f]{Huangyu~Xiao\,\orcidlink{0000-0003-2485-5700}}
\affiliation[a]{Physics Department, Indiana University, Bloomington, IN 47405, USA}
\affiliation[b]{Department of Physics, Enrico Fermi Institute, Leinweber Institute for Theoretical Physics, and Kavli Institute for Cosmological Physics, University of Chicago, Chicago, IL 60637, USA}
\affiliation[c]{Kavli Institute for the Physics and Mathematics of the Universe (WPI),\\
    The University of Tokyo Institutes for Advanced Study,\\
    The University of Tokyo, Kashiwa, Chiba 277-8583, Japan}
\affiliation[d]{Theoretical Physics Division, Fermi National Accelerator Laboratory, Batavia, IL, USA}
\affiliation[e]{Physics Department, Boston University, Boston, MA 02215, USA}
\affiliation[f]{Department of Physics, Harvard University, Cambridge, MA, 02138, USA}
\emailAdd{rco@iu.edu}
\emailAdd{kharigaya@uchicago.edu}
\emailAdd{isaacw@fnal.gov}
\emailAdd{hxiao3@bu.edu}

\preprint{CETUP2025-015, FERMILAB-PUB-25-0812-T}

\abstract{
We introduce a novel mechanism where the kinetic energy of a rotating axion can be dissipated by the interactions with dark magnetic monopoles. This mechanism leads to a framework where the QCD axion and dark monopoles account for the dark matter density, and the observed baryon asymmetry is generated through the rotating QCD axion via axiogenesis.
The monopoles acquire masses from a nonzero axion field, and they can transition between different quantized dyonic levels in the presence of a rotating axion field.
The axion kinetic energy is dissipated by the transition, and thus the axion abundance is depleted to the observed dark matter abundance. We predict that the axion decay constant should be below $10^9$ GeV to explain the observed dark matter and baryon densities.
}

\begin{document}
\maketitle
\flushbottom

\section{Introduction}
The magnetic monopole is a theoretically well-motivated object. It is a generic consequence of spontaneous symmetry breaking of gauge theories when the vacuum manifold has nontrivial topology.
In the simplest UV completion, an $SU(2)$ gauge symmetry is broken to $U(1)$ and a ’t Hooft–Polyakov monopole arises as a smooth and finite-energy soliton~\cite{Polyakov:1974ek,tHooft:1974kcl}. 
The cosmological abundance of monopoles is expected to be generated during the phase transition when gauge symmetry is spontaneously broken. As stable particles carrying magnetic charges, they overclose the Universe if the symmetry breaking scale is around the typical grand unification scale, known as the monopole problem~\cite{Zeldovich:1978wj,Albrecht:1982wi,Preskill:1984gd}.
If one considers magnetic monopoles originating from a dark sector, the monopole mass can be significantly lower than that considered in grand unified theories, and its abundance determined by the Kibble-Zurek mechanism and freeze-out mechanism naturally accounts for the dark matter relic abundance provided that the dark monopole mass is above 100 TeV~\cite{Preskill:1979zi,Murayama:2009nj}.

\begin{figure}[t!]
    \centering
    \includegraphics[width=0.6\linewidth]{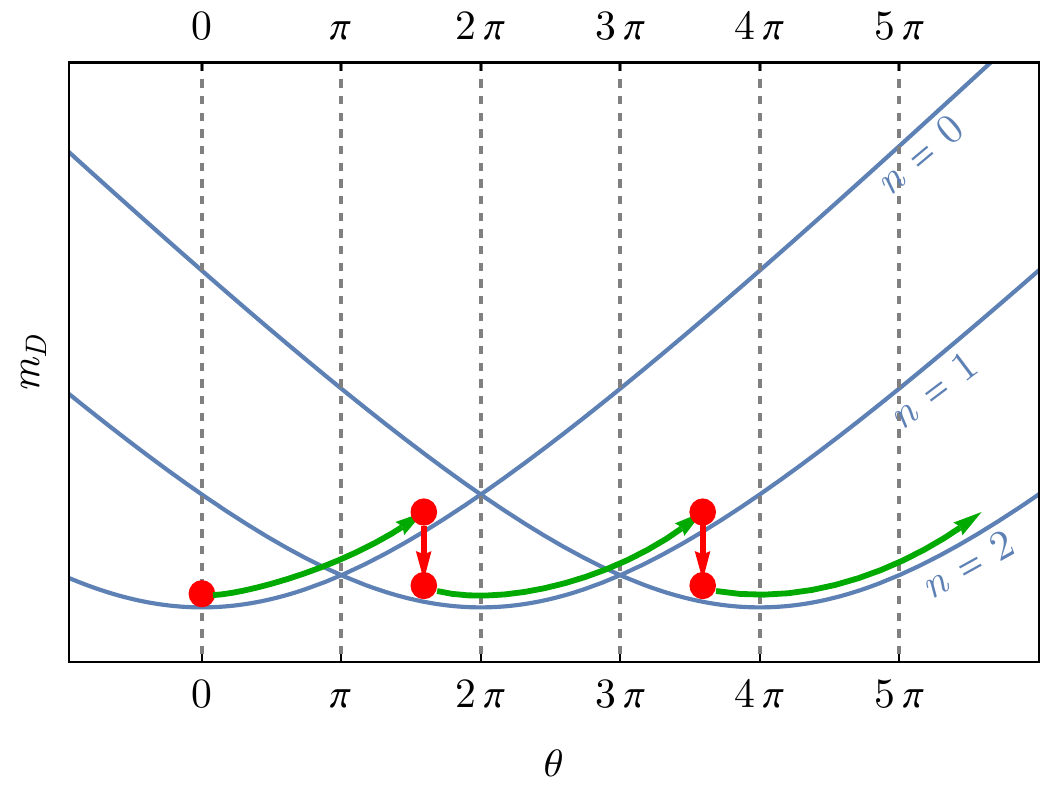}
    \caption{Cartoon for the axion kinetic energy dissipation process with $\theta$ the axion misalignment angle and $m_D$ the dyon mass.
    The dissipation is achieved through the climbing of the axion field over many cycles, which is only enabled with the level crossing effect. During the rotation of the axion field, the axion kinetic energy is converted to potential energy. During the level crossing, the axion kinetic energy is conserved, but the potential energy is converted to a pair of dark fermions that the dyon decays into. 
}
    \label{fig:cartoon}
\end{figure}

The dynamics of monopoles is also rich, which has been extensively studied in monopole-fermion interactions~\cite{Rubakov:1981rg,Callan:1982ac,Csaki:2021ozp,Brennan:2021ewu,Brennan:2023tae,Brennan:2024sth,vanBeest:2023mbs,Bogojevic:2024xtx,Csaki:2024ajo,Loladze:2024ayk,Khoze:2024hlb,Dawson:1983cm,Arafune:1983uz} and monopole-axion interactions~\cite{Kawasaki:2015lpf,Nomura:2015xil,Kawasaki:2017xwt,Banerjee:2024ykz,Fan:2021ntg,GarciaGarcia:2025uub}. As a specific example, the monopole can exhibit interesting behavior
in the presence of a $\theta$ term, since the monopole can acquire an electric charge via the Witten effect~\cite{Witten:1979ey}. When $\theta$ is dynamical, the electrostatic contribution to the monopole mass varies cosmologically, and the effective axion potential also receives contributions from monopoles~\cite{Fan:2021ntg,GarciaGarcia:2025uub}. 

In this work, we point out a unique and interesting phenomenon in which the monopoles can transition between different quantized dyonic levels when the axion field is rotating in its field space, as shown in Fig.~\ref{fig:cartoon}. A rotating axion field is a consequence of explicit Peccei-Quinn (PQ) symmetry breaking and can generate baryon asymmetry via the axiogenesis mechanism~\cite{Co:2019wyp,Domcke:2020kcp,Co:2020xlh}. During times when $\dot{\theta}\equiv \dot{a}/f_a \neq 0$, adjacent electric levels on a magnetic monopole may cross by releasing energy and transferring electric charge to light dark fermions that are charged under the same gauge group as the dark monopoles. A related study can be found in Ref.~\cite{Jeong:2023gjc}, which also uses monopole-axion dynamics to dissipate the energy of axion oscillations and modify axion abundance from the misalignment mechanism.%
\footnote{Ref.~\cite{Jeong:2023gjc} studies the dissipation by the emission of charged vector bosons, although it is not clear if such emission is kinematically possible. Also, as noted in the paper, the level crossing by axion oscillations requires an axion-gauge field coupling much larger than the natural strength shown in Eq.~\eqref{eq:axion coupling}.}
In contrast, our study focuses on the dissipation of axion rotation, which serves as the source of the baryon asymmetry as well as part of the dark matter from kinetic misalignment.

The cosmological history of the axion rotation in our scenario is as follows. 
Axion rotation is initiated by the Affleck-Dine mechanism~\cite{Affleck:1984fy}, and the observed baryon asymmetry is produced by axiogenesis around the electroweak phase transition.
The axion-driven level crossing of monopoles dissipates the kinetic energy of the rotating QCD axion, which would otherwise overproduce axion dark matter via kinetic misalignment~\cite{Co:2019jts,Eroncel:2022vjg,Fasiello:2025ptb}.%
The residual kinetic energy produces axion dark matter.
See Refs.~\cite{Co:2021rhi,Madge:2021abk} for a possible dissipation mechanism of axion rotation by the production of a helical Abelian gauge field, although the helicity of the gauge fields may convert back to the axion rotation, leaving a significant amount of axion rotation. See~\cite{Co:2021rhi} for a free-energy argument favoring the suppression of the axion rotation.

Our scenario addresses three major problems of the Standard Model: the strong CP problem is solved by the QCD axion through the PQ mechanism~\cite{Peccei:1977hh,Peccei:1977ur}, the baryon asymmetry is generated by axiogenesis, and dark matter is composed of dark monopoles, dark fermions, and QCD axions.

This paper is organized as follows.
In Sec.~\ref{sec:setup}, we define our model and review how baryogenesis is achieved via axiogenesis.
In Sec.~\ref{sec:dissipation}, we introduce our mechanism in detail by calculating the axion rotation dissipation rate and solving the evolution of the coherent rotation.
In Sec.~\ref{sec:DM}, we calculate the dark matter relic abundance from axions and dark fermions.
In Sec.~\ref{sec:conclusion}, we summarize our findings and discuss the implications of our mechanism, highlighting its predictions for axion searches and dark-matter phenomenology.

\section{Model Setup}
\label{sec:setup}

\subsection{Dark Monopoles}
We consider a dark gauge sector in which an $SU(2)_D$ gauge symmetry is spontaneously broken to a $U(1)_D$ by the vacuum expectation value (VEV) $v_D$ of an adjoint Higgs field. This allows for 't Hooft–Polyakov monopoles because $\pi_2(SU(2)_D/U(1)_D)=\mathbb{Z}$, providing the simplest UV completion that yields magnetic charge in the infrared. The massive vector bosons have a mass
$m_W = e_D\,v_D$, where $e_D$ is the $U(1)_D$ gauge coupling.
The monopole magnetic charge is $Q_M=4\pi/e_D$.
We will consider a monopole mass around 100--500 TeV, which allows for stable monopoles constituting a fraction $f_M$ of the dark matter (DM) energy density today, determined by the Kibble-Zurek and freeze-out mechanisms~\cite{Preskill:1979zi,Murayama:2009nj}.
Light dark fermions $f$, neutral under the Standard Model (SM) gauge group but charged under $U(1)_D$, will play a key role in enabling dyon level crossings. If $f$ originates from an $SU(2)_D$ doublet, its $U(1)_D$ charge is $\pm e_D/2$.
More specifically, we consider two $SU(2)$ doublets, $(f_1,f_2^c)$, and $(f_2,f_1^c)$, as the simplest anomaly-free theory.

The QCD axion $a$ is assumed to generate the observed baryon asymmetry via axiogenesis as discussed below, while also coupling anomalously to the dark gauge sector. At energies below the $SU(2)_D$ breaking scale, this coupling takes the form
\begin{equation}
\label{eq:axion coupling}
\mathcal{L}\supset \frac{a}{f_a}\,\frac{\alpha_D}{8\pi} F'_{\mu\nu}\tilde F'^{\mu\nu},
\end{equation}
with $F'$ the $U(1)_D$ field strength and $\alpha_D\equiv e_D^2/4\pi$. This coupling promotes the dark vacuum angle to the axion field value, $\theta\equiv a/f_a$, leading to an axion-dependent Witten effect for monopoles and dyons.

In the presence of the QCD axion background, the monopole acquires extra charges and masses, becoming a dyon. The QCD axion mass also receives corrections from the instanton in the dark sector, $e^{-S_{\rm inst}}=e^{-2\pi/\alpha_D}$, which is exponentially suppressed for our choice of parameters.
The strong CP problem thus remains solved by the QCD axion.
In the BPS limit with electric level $n\in\mathbb{Z}$, the dyon mass is $m_D^2 = m_M^2 + m_W^2\left(n-{\theta}/{2\pi}\right)^2$, where $m_M=4\pi v_D/e_D$ is the mass of the purely magnetic monopole. Before introducing light dark fermions, the dyon is stable against losing electric charge
by emitting charged vector bosons~\cite{Weinberg:2012pjx}\footnote{In the non-BPS limit, the emission of charged vector bosons may be possible. Still, the Universe ends up with too many of these bosons, which leads to overclosure~\cite{Brummer:2025inh}.}. 
After light fermions with mass $m_f$ are introduced to the theory, the charge radius of the dyon parametrically increases from $1/m_M$ to $1/m_f$~\cite{Hook:2024als,Grossman:1983yf,Yamagishi:1982wp}, thereby lowering the electrostatic contribution to the dyon mass.
The dyon mass as a function of the $\theta$ value has been calculated as~\cite{Hook:2024vhf}
\begin{equation}\label{eq:dyon_mass}
m_D \approx 2\,m_f \,{\rm sin}^2\left(\frac{\pi}{2}\left[n-\frac{\theta}{2\pi}\right]\right)+\frac{\alpha_D}{2}m_f\left(n-\frac{\theta}{2\pi}\right)^2\,.
\end{equation}
The difference of the dyon masses between the adjacent levels, though lowered by the light fermions, can be larger than $2 m_f$ when $\theta$ approaches $2\pi$~\cite{Hook:2024vhf}, allowing for the dyon to decay to a pair of fermions with certain $\theta$ values per period.%
\footnote{
 The fermion dynamics reduce to a 1+1-dimensional Dirac equation~\cite{Kazama:1976fm,Grossman:1983yf,Yamagishi:1982wp,Hook:2024vhf}, where pair production is fixed by the axial anomaly and is unsuppressed. Solving the Dirac equation gives a rate $\Gamma = (v - 2m_f)/\pi$ when $v > 2m_f$, which for $v \gtrsim 2m_f$ implies $\Gamma = \mathcal{O}(m_f)$~\cite{Blaer:1981ps}. We find such a rate is much larger than the Hubble rate in our parameter space when the decay kinematically open and will thus treat the decay as instantaneous.
}
Since the dark fermions are $SU(2)$ doublets, the two components will have different charges, with $f_1$ charge $+e_D/2$ and $f_2$ charge $-e_D/2$.
The pair production process is $M_{n-1}\rightarrow M_n+{f_1^c}+f_2$, where $n$ denotes the quantized dyonic level, and both outgoing particles carry charge $-e_D/2$.

As $\theta$ evolves due to the axion rotation, the charge of dyons changes dynamically, enabling level crossings and releasing the corresponding mass-gap energy.
Changes in $\theta$ are continuously accompanied by transitions between dyonic levels $n$.
Fermion emissions by dyons were studied in Refs.~\cite{Blaer:1981ps,Blaer:1981ui,Marciano:1983pu,Lam:1984sm} and more recently in Ref.~\cite{Hook:2024vhf} with a static $\theta$. 
Our scenario instead uses a dynamical axion field to induce level crossings of charge-varying dyons.%

\subsection{Axiogenesis}
Baryogenesis in this work is accomplished via the axiogenesis mechanism~\cite{Co:2019wyp,Domcke:2020kcp,Co:2020xlh}.
A conserved charge, known as the PQ charge or asymmetry, can be carried by rotation in the angular (axion) direction of field space. The charge density is given by $n_{\rm \theta} = \dot{\theta} f_a^2$. Analogous to the dynamics in the Affleck-Dine mechanism~\cite{Affleck:1984fy}, this charge, or rotation, can be initiated by explicit PQ breaking from higher-dimensional operators that may be expected when the PQ symmetry is an accidental symmetry or from quantum gravity effects on any global symmetry.
The rotation is initially elliptical, but by the interaction with the thermal bath, the motion takes a form that minimizes the free energy for a fixed PQ charge, namely, circular motion~\cite{Co:2019wyp,Domcke:2022wpb}. The radius of the rotation shrinks due to cosmic expansion until it reaches its minimum.

In the minimal realization~\cite{Co:2019wyp}, the PQ charge is transferred into the chiral asymmetry of SM fermions via the QCD sphaleron process and subsequently into baryon asymmetry by electroweak sphalerons.
The resultant baryon asymmetry is
\begin{align}
    Y_B = \frac{45}{2 \pi^2 g_*} \frac{c_B \dot{\theta}_{\rm EW}}{T_{\rm EW}}\,,
\end{align}
where $T_{\rm EW} \simeq 130$ GeV and $\dot{\theta}_{\rm EW}$ are the temperature and angular velocity when the electroweak sphaleron decouples.
$c_B$ is a constant determined by solving the transport equations in the thermal bath for the charge transfer.
In the minimal scenario with the SM extended by an axion, $c_B \simeq 0.1$. The observed value of the baryon asymmetry then fixes the amount of the PQ charge by
\begin{align}
    \label{eq:YB from Ytheta}
    Y_B = c_B \frac{T_{\rm EW}^2}{f_a^2} Y_\theta\,.
\end{align}

The rotating axion field further evolves and finally constitutes axion dark matter.
$Y_\theta$ remains constant until the axion rotation gets trapped by the potential energy or fragments into axion fluctuations, after which the axion starts oscillating around the bottom of the potential, known as kinetic misalignment~\cite{Co:2019jts,Co:2020dya,Eroncel:2022vjg,Fasiello:2025ptb}.
In the minimal scenario, the dark matter is overproduced by a factor of 70 for the $Y_\theta$ that produces the observed baryon number asymmetry.
Previous works have aimed to increase the efficiency of baryon number production to resolve this tension~\cite{Co:2020jtv,Harigaya:2021txz,Chakraborty:2021fkp,Kawamura:2021xpu,Co:2021qgl,Barnes:2022ren,Co:2022kul,Barnes:2024jap}.
In this work, we will show that the axion DM abundance can be depleted through the axion-monopole interactions, reducing it significantly compared to the minimal scenario and thereby resolving the tension.

\section{Axion Energy Dissipation} 
\label{sec:dissipation}

\subsection{Kinetic Energy Dissipation}
We now discuss the dissipation rate of the coherent axion rotation.
A level crossing occurs once every $2 \pi$, each time releasing an energy of $2 m_f$, and thus the energy dissipation rate of the axion is given by
\begin{align}
    \dot{\rho}_\theta = \frac{\dot{\theta}}{2 \pi} n_M \cdot (2m_f) =  \frac{\dot{\theta}}{\pi} \rho_M \frac{m_f}{m_M} =  \frac{\dot{\theta}}{\pi} \rho_M \alpha_D \frac{m_f}{m_W} \,.
\end{align}
Here, $n_M$ and $\rho_M$ are the number density and energy density of the monopole, respectively.
If the axion coupling to the dark $U(1)$ in Eq.~\eqref{eq:axion coupling} is enhanced, the dissipation rate will increase correspondingly.
This rate should be normalized by the axion energy density itself to know whether the axion energy density can be fully dissipated.
To do this, one can recall that the axion rotational energy density is given by
\begin{align}
    \label{eq:rho_theta}
    \rho_\theta = \frac{1}{2} \dot{\theta}^2 f_a^2 = \frac{1}{2} \dot{\theta} n_\theta \, .
\end{align}
Using the definition of $n_\theta \equiv s Y_\theta$, one arrives at the dissipation rate due to level crossing of monopoles
\begin{align}
\label{eq:Gamma_theta}
    \Gamma_\theta \equiv \frac{\dot{\rho}_\theta}{\rho_\theta} = \frac{2\alpha_D}{\pi} \frac{m_f}{m_W} \frac{1}{Y_\theta} \frac{\rho_M}{s} = \frac{2\alpha_D}{\pi} \frac{m_f}{m_W}\frac{f_M \xi_{\rm DM}}{Y_\theta},
\end{align}
where $s$ is the entropy density, and $f_M \equiv \rho_M/\rho_{\rm DM}$ is the fraction that monopole contributes to the DM.
Here we denote the redshift-invariant, observed dark matter abundance by $\xi_{\rm DM} \equiv \rho_{\rm DM}/s = 0.44~\rm eV$.
For convenience, we define $\epsilon \equiv 2 \alpha_D m_f f_M /\pi m_W$ to write
\begin{align}
    \label{eq:dissipation rate}
    \Gamma_\theta = \epsilon \frac{\xi_{\rm DM}}{Y_\theta}  \, .
\end{align}

With dissipation, the PQ charge yield is no longer a constant but decreases over time.
In the presence of cosmic expansion, Eq.~(\ref{eq:Gamma_theta}) generalizes to $\dot Y_\theta = \Gamma_\theta Y_\theta/2$, which, together with Eq.~\eqref{eq:dissipation rate}, leads to
\begin{align}
\label{eq:Y with time}
    Y_\theta(t) = Y_\theta - \frac{\epsilon \xi_{\rm DM}}{2} (t - t_{\rm EW})\,,
\end{align}
where the initial value $Y_\theta$ is given by Eq.~\eqref{eq:YB from Ytheta} at the time of the electroweak phase transition.
We can safely ignore $t_{\rm EW}$ as the dissipation becomes effective much later.
From Eq.~\eqref{eq:Y with time}, one immediately finds
\begin{equation}
    \dot{\theta}(t)  = \frac{s(t)Y_\theta(t)}{f_a^2}   = (2 b M_{\rm Pl}^3)^{1/2} \left( \frac{Y_B}{c_B T_{\rm EW}^2 t^{3/2}} - \frac{\epsilon \xi_{\rm DM}}{2 f_a^2 \sqrt{t}}\right) \, ,
\end{equation}
where we have translated the temperature $T$ to time $t$ in the expression of entropy.
The relationship between $T$ and $t$ can be derived by matching the expression of the Hubble rate $H(T) = b  T^2/M_{\rm Pl} = H(t) =  1/(2t)$, where we define $b = (\pi^2 g_*/90)^{1/2}$ and $M_{\rm Pl} = 2.4 \times 10^{18}$~GeV is the reduced Planck mass.

Dissipation becomes efficient when the above rate is comparable to three times the Hubble rate.
This gives the dissipation temperature
\begin{align}
    \label{eq:dissipation T}
    T_{\rm di} &= \frac{1}{2}\left( \frac{7\epsilon}{3 b Y_\theta} \right)^{1/2} (\xi_{\rm DM} M_{\rm Pl})^{1/2} \,.
\end{align}
Since the effective number of relativistic degrees of freedom, $g_*$, is not sensitive to temperature in the range considered in this work, we fix $g_* = 70$.
Using Eq.~\eqref{eq:YB from Ytheta}, we obtain the dissipation temperature
\begin{align}
    \label{eq:Td with YB}
    T_{\rm di} =&\ \frac{1}{2} \left( \frac{7\epsilon}{3b} \right)^{1/2} \left( \frac{c_B}{Y_B} \right)^{1/2} \frac{T_{\rm EW}}{f_a} \left( \xi_{\rm DM} M_{\rm Pl} \right)^{1/2} \\
           \simeq & \ 3~\gev \left(\frac{\alpha_D}{0.2}\right)^{\frac{1}{2}} \left(\frac{f_M}{0.5}\right)^{\frac{1}{2}}  \left(\frac{m_f}{0.01 m_W}\right)^{\frac{1}{2}}  \left(\frac{10^9~\gev}{f_a}\right) \left(\frac{c_B}{0.3}\right)^{\frac{1}{2}} \nonumber \,.
\end{align}
Once dissipation begins, the zero-mode coherent rotation quickly diminishes until it is trapped by the axion mass or destroyed by backreaction from nonzero modes generated via parametric resonance.
For successful axiogenesis, $T_{\rm di}$ must be below $T_{\rm EW}$, which gives a weak constraint
\begin{equation}
    f_a \gtrsim 9 \times  10^6 ~\mathrm{GeV} \left( \frac{\alpha_D}{0.1} \right)^{1/2} \left(\frac{f_M}{0.5}\right)^{1/2} 
    \left( \frac{m_f}{0.01 m_W} \right)^{1/2} \left( \frac{c_B}{0.1} \right)^{1/2}\,.
\end{equation}

\subsection{Gradient Energy Dissipation}
\label{subsec:nonzero_di}
Besides the zero mode, the nonzero modes may also be dissipated. Nonzero modes, i.e., fluctuations, arise because the rotation of the complex scalar field spontaneously breaks both the global $U(1)$ and time-translational symmetries, leading to a gapless, phonon mode corresponding to fluctuations of the PQ charge density that are excited by cosmic (curvature) perturbations. The production of these fluctuations are discussed in Refs.~\cite{Eroncel:2025bcb,Eroncel:2025qlk,Bodas:2025eca} and will be referred to as the acoustic misalignment mechanism (AMM) production.%
\footnote{Before the rotation is thermalized, the fluctuations around elliptical rotation can grow by parametric resonance~\cite{Co:2017mop,Co:2020dya,Co:2020jtv} and contribute to dark matter. We assume that the thermalization occurs before parametric resonance becomes effective.}

The nonzero dissipation rate depends on how $k \delta \theta$, with $\delta \theta$ the amplitude of the fluctuation, compares to the zero mode $\dot{\theta}$ and the fermion mass $m_f$.
The comparison depends on the specific wavenumber $k$.
Before categorizing different $k$ modes, we discuss the condition on $k \delta \theta$ for efficient dissipation of the nonzero modes.
In the case where the zero-mode energy density is negligible compared to that of the fluctuations, we can estimate the rate using 
\begin{align}
    \dot{\rho}_k = \frac{|\dot{\theta}|}{2 \pi} n_M \cdot (2m_f) \simeq \frac{2 k \delta\theta m_f n_M}{\pi^2}  \, ,
\end{align}
where $|\dot{\theta}| \simeq |k\delta\theta \sin(k t)|$ averages to $2k\delta\theta/\pi$ over one oscillation period. This gives the rate
\begin{align}
\label{eq:rate fluc}
    \Gamma_k \equiv \frac{\dot{\rho}_k}{\rho_k} = \frac{2 m_f n_M}{\pi^2 \delta\theta k f_a^2}.
\end{align}

Even when the zero-mode motion exists, as long as $k \delta\theta > \dot\theta$,
the frequency of level crossings depends on the fluctuation amplitude, and their dissipation proceeds at the rate given in Eq.~\eqref{eq:rate fluc}.
If $k \delta\theta < \dot\theta$, on the other hand, the fluctuations no longer affect the frequency of level crossings and are therefore not dissipated. 
However, as the dissipation of the zero mode proceeds,  $\dot\theta$ becomes smaller than $k \delta\theta$ and the dissipation of the fluctuations turns on.

Note that dissipation can only proceed when the mass gap is larger than the kinematic threshold $\left|m_D(n+1)-m_D(n)\right|_{\theta=\pi+\delta}>m_f$. Using the dyon mass expressions from Refs.~\cite{Hook:2024als,Grossman:1983yf,Yamagishi:1982wp}, we find that $\delta\theta$ must exceed $O(1)$ for the nonzero modes to be dissipated. The fluctuations are continuously dissipated until $\delta \theta$ drops to $O(1)$.

\subsection{Washout by $SU(2)_D$ Sphaleron}
Before the $SU(2)_D$ phase transition, $SU(2)_D$ sphaleron processes are active and can dissipate the axion kinetic energy since the PQ symmetry has an $SU(2)_D$ anomaly.
However, in the limit $m_f=0$, a linear combination of the PQ symmetry and the chiral symmetry of the fermion does not have an $SU(2)_D$ anomaly, and thus dissipation cannot occur.
Dissipation therefore requires both the $SU(2)_D$ sphaleron process and chiral symmetry breaking, with the rate given by~\cite{McLerran:1990de,Co:2019wyp}
\begin{align}
    \Gamma_{SU(2)_D}  \simeq {\rm min}\left( \alpha_D \frac{m_f^2}{T}, 120\alpha_D^5 T \right) \times \frac{T^2}{S^2},
\end{align}
where $S$ denotes the radius of the axion rotation, which exceeds $f_a$ at high temperatures,
\begin{align}
S^2 = f_a^2\times{\rm max}\left(1, \left(\frac{T}{T_S}\right)^k  \right),
\end{align}
where $k=2$ ($k=3$) when the potential of the radial direction of the PQ breaking field is quartic (quadratic) at $S\gg f_a$.
In the orange-shaded region in Fig.~\ref{fig:main}, the $SU(2)_D$ sphaleron process washes out the axion rotation before the electroweak phase transition for $k=3$, i.e., $\Gamma_{SU(2)_D} > 3 H$ at $T_{\rm EW}$, where we have assumed $m_S = 10~\rm MeV$.
Other constraints in the figure will be discussed in Sec.~\ref{sec:DM}.

Although we focus on the case where axion rotation is dissipated via monopoles, we point out that it is still possible to dissipate the axion rotation by the $SU(2)_D$ sphaleron process after the electroweak phase transition for certain choices of $m_W$ and $m_f$.
This requires $m_W$ to be below the electroweak scale, in which case the monopole dark matter abundance becomes negligible.
Entropy in the $SU(2)_D$-charged sector can be transferred to the SM sector through a quartic coupling between the SM Higgs and the $SU(2)_D$-breaking Higgs field, so that dark radiation is not overproduced.
We leave a detailed study of this case and its potential signatures, such as rare Higgs boson decays, for future work.

\section{Dark Matter}
\label{sec:DM}
With the presence of monopoles, dark fermions, and the axion, dark matter now comprises multiple components.
In this section, we discuss the relic abundance for these components.

\subsection{Axion Dark Matter}
\begin{figure}[t!]
    \centering
    \includegraphics[width=0.7\linewidth]{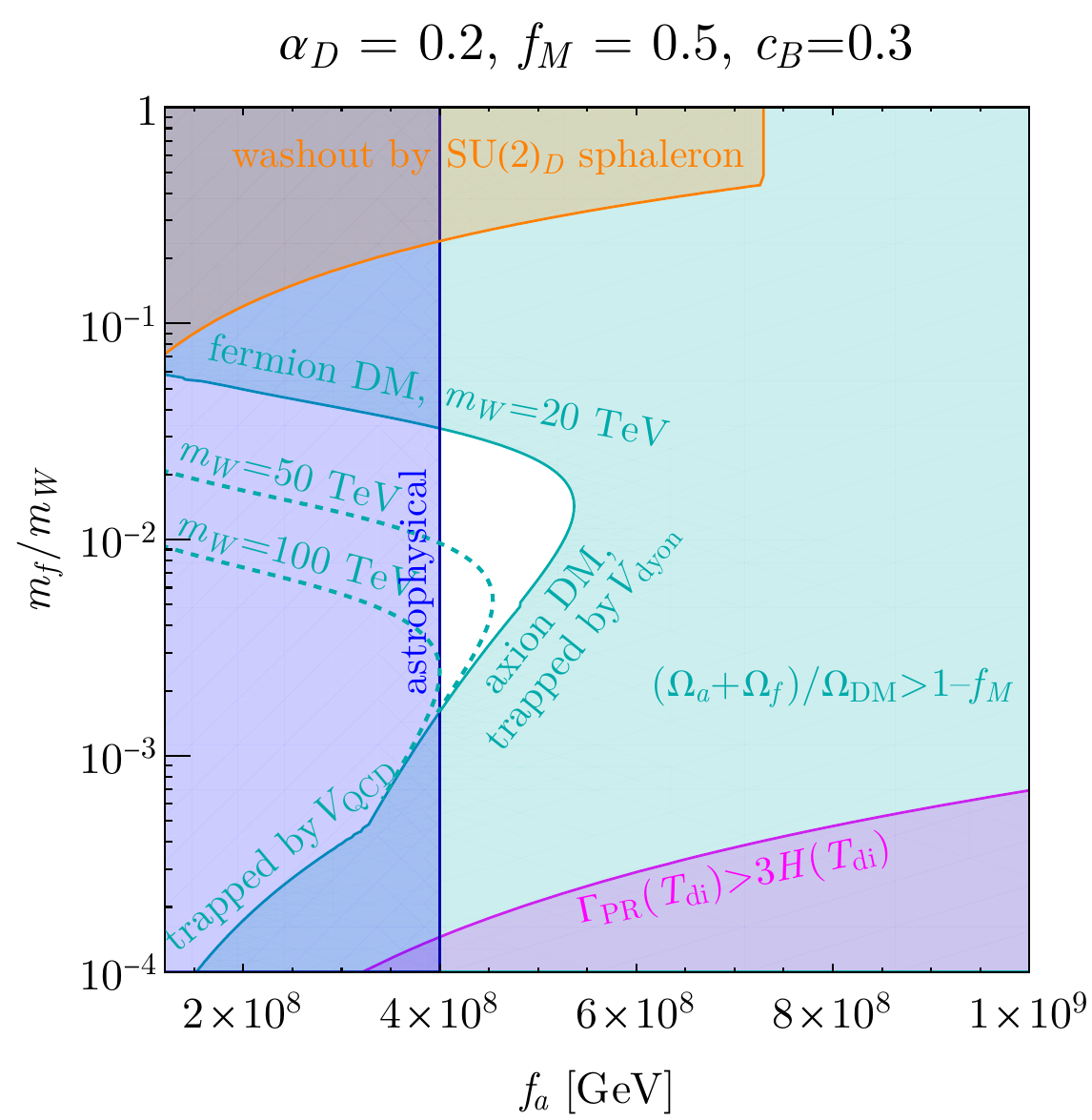}
    \caption{Parameter space for successful baryogenesis and dark matter with fixed $\alpha_D = 0.2$, $c_B = 0.3$, and $f_M = 0.5$.  Pink: PR becomes effective before $T_{\rm di}$ so that the coherent zero mode is converted into fluctuations and cannot be dissipated. Green: with monopole accounting for $f_M=50\%$ of dark matter, dark matter is overproduced by the corresponding particle labeled along the boundary. We show contours of $m_W$ necessary to help account for 100\% dark matter. Orange: axion rotation is washed out by the $SU(2)_D$ sphaleron process before EWPT, for $m_S = 10~\rm MeV$, so that axiogenesis is not achieved. Blue: excluded by astrophysical bounds assuming the KSVZ axion.
    Note that the astrophysical bound may be relaxed~\cite{DiLuzio:2017ogq,Bjorkeroth:2019jtx,Badziak:2023fsc}.
    }
    \label{fig:main}
\end{figure}

We start by discussing the contribution due to the axion zero mode.
The final axion relic density is determined when the zero mode is trapped by the QCD axion potential or the dyon-axion potential
\begin{equation}
\label{eq:dyon-axion}
    V_{\rm dyon-axion} = m_D(a) n_M .
\end{equation}
The zero mode is trapped when the kinetic energy is no longer larger than the potential barrier,
\begin{align}
\frac{1}{2}\dot{\theta}^2(t_{\rm trap}) f_a^2 = \max \left[
    2m_a^2(t_{\rm trap}) f_a^2\,,
    2m_f n_M(t_{\rm trap}) \,\right]\,,
\end{align}
where the first expression on the right is the QCD potential barrier, while the second one is the dyon-axion potential barrier.
Here, $m_a(t)$ is the time-dependent QCD axion mass, typically written as a function of temperature as
\begin{align}
m_a(T) &= m_{a0} \times\begin{cases}
    \left(\Lambda_{\rm QCD}/T\right)^4\,, &T \gtrsim \Lambda_{\rm QCD}\,,\\
    1,& T \lesssim \Lambda_{\rm QCD}\,.
\end{cases}\,, \\
m_{a0} & \simeq 6~\mathrm{meV} \left(\frac{10^9~\gev}{f_a}\right)\,,
\end{align}
where $\Lambda_{\rm QCD} \simeq 146~\rm MeV$ is the QCD scale and $m_{a0}$ is the QCD axion mass at low temperature.
The axion number density at this moment is given by%
\footnote{
After trapping, the oscillation of the axion around the minimum with $m_a \ll H$ can create fluctuations via parametric resonance~\cite{Arvanitaki:2019rax,Harigaya:2025pox}. The self-scattering of the axion fluctuations can reduce their number density by an $O(1)$ factor~\cite{Co:2025jnj}.
}
\begin{align}
    n_a(t_{\rm trap}) \simeq \dot{\theta}(t_{\rm trap}) f_a^2\,.
\end{align}
The axion relic abundance can then be computed as $\rho_a = m_{a0} n_a(t_{\rm trap})/s(t_{\rm trap})$. The final results read
\begin{align}
\label{eq:axion DM}
    \frac{\rho_a}{s}  \simeq \xi_{\rm DM} \times \begin{cases}
    \left( \frac{f_a}{6\times10^8 ~{\rm GeV}} \right)^7 
    \left( \frac{10^{-4}}{\epsilon} \right)^{\frac{7}{2}}
    \left( \frac{0.25}{c_B} \right)^{\frac{7}{2}}  \\ 
    \left( \frac{f_a}{6\times10^8 ~{\rm GeV}} \right)^{\frac{3}{2}} 
    \left( \frac{10^{-4}}{\epsilon} \right)^{\frac{1}{4}}
    \left( \frac{0.25
    }{c_B} \right)^{\frac{3}{4}} ,
    \end{cases}
\end{align}
with the upper (lower) expression for trapping by the QCD (dyon) potential. These analytic expressions give good approximations in the region of parameter space in Fig.~\ref{fig:main}.

The coherent axion rotation can be destroyed by parametric resonance~\cite{Kofman:1994rk,Kofman:1997yn}, and dissipation is no longer effective before the axion gets trapped.
This scenario is known as axion fragmentation~\cite{Fonseca:2019ypl,Eroncel:2022vjg} (see also~\cite{Jaeckel:2016qjp,Berges:2019dgr,Co:2021rhi}), with the rate given by
\begin{equation}
    \Gamma_{\rm AF} = \frac{m_a^2}{2\dot\theta}\,,
\end{equation}
at the resonance peak with $k = \dot{\theta}/2$ and width $\Delta k = m_a^2/\dot{\theta}$.
For a certain moment of time, the effective production rate is thus given by
\begin{align}
\label{eq:PR rate}
    \Gamma_{\rm PR} = \Gamma_{\rm AF} \frac{\Delta k}{k} = \frac{m_a^4}{\dot{\theta}^3}\,.
\end{align}
Specifically, the axion-dyon potential can also contribute to PR.
The corresponding rate can be calculated by replacing $m_a^2$ with $m_f n_M/f_a^2$ in Eq.~\eqref{eq:PR rate} according to Eq.~(\ref{eq:dyon-axion}) with the approximation that $m_D \simeq m_f$.

For successful dissipation, PR production cannot happen before $T_{\rm di}$.
The rotation speed at $T_{\rm di}$ can be calculated as
\begin{equation}
    \dot{\theta}_d  = \frac{s Y_\theta (T_{\rm di})}{f_a^2}
         = \frac{2}{9}\sqrt{21b} \sqrt{\frac{c_B}{Y_B}} \epsilon^{3/2} \frac{\left(\xi_{\rm DM} M_{\rm{Pl}}\right)^{3/2}}{f_a^3} T_{\rm EW}\,,
\end{equation}
where we have used Eq.~\eqref{eq:Td with YB}.
Requiring that $\Gamma_{\rm PR}(T_{\rm di}) < 3 H(T_{\rm di})$, we obtain
\begin{equation}
    \label{eq:Gamma_PR bound}
    f_a \lesssim 10^{12}~\gev \left(\frac{c_B}{0.1}\right)^{21/46}  \left(\frac{\alpha_D}{0.1}\right)^{27/46}  \left(\frac{m_f}{0.01 m_W}\right)^{27/46} f_M^{27/46}\,,
\end{equation}
for PR driven by the QCD axion potential, and
\begin{align}
    f_a \lesssim & ~ 10^{14}~\gev \left(\frac{0.1}{c_B}\right)^{\frac{1}{2}} 
    \left(\frac{\alpha_D}{0.1}\right)^{\frac{1}{2}} \left(\frac{f_M}{0.5}\right)^{\frac{1}{2}} \left(\frac{m_f}{0.01 m_W}\right)^{\frac{1}{2}} \,.
\end{align}
for PR driven by the dyon-axion potential.

The time $t_{\rm PR}$ when backreaction destroys the coherent rotation can be solved by comparing the energy densities of the zero mode and the nonzero mode.
The energy density for a nonzero mode with a certain momentum $k$ is given by
\begin{align}
    \rho_k(t) = \frac{1}{2} \delta \theta (t)^2 k^2 f_a^2\,.
\end{align}
The total energy across all $k$ can be estimated as
\begin{align}
    \rho_{\rm PR} \simeq \frac{1}{2} \delta \theta_0^2 \left(\frac{\dot{\theta}}{2}\right)^2 f_a^2 e^{\Gamma_{\rm PR}/\mathrm{max}(3H,\Gamma_\theta)}\,,
\end{align}
where we have used $k \simeq \dot{\theta}/2$, and
$\delta \theta_0 \simeq \sqrt{P_\zeta}$ is the initial value of the fluctuation. Here $P_\zeta \simeq 2 \times 10^{-9}$~\cite{Planck:2018vyg} is the adiabatic perturbation.
$t_{\rm PR}$ then can be derived by requiring $\rho_{\theta} = \rho_{\rm PR}$, with $\rho_{\theta}$ given by Eq.~\eqref{eq:rho_theta}.
The yield of axion can then be computed at $t_{\rm PR}$ as
\begin{align}
    n_a(t_{\rm PR}) \simeq \dot{\theta}(t_{\rm PR}) f_a^2\,.
\end{align}
The axion relic abundance can then be computed as $\rho_a = m_{a0} n_a(t_{\rm PR})/s(t_{\rm PR}) $.

Fig.~\ref{fig:main} shows the parameter space of our mechanism in the $f_a$--$m_f/m_W$ plane.
The total DM is overproduced in the cyan shaded region, with the fermion relic abundance discussed in the next section.
We fix $c_B = 0.3$, $\alpha_D = 0.2$, and $f_M = 0.5$ as an optimized choice.
The axion DM relic abundance provides an upper bound on $f_a$, corresponding to the cyan-shaded exclusion region on the right side of the available parameter space.
We label the boundaries corresponding to the trapping by the QCD potential and axion-dyon potential in Fig.~\ref{fig:main}.
The condition Eq.~\eqref{eq:Gamma_PR bound}, on the other hand, is always covered by the DM production requirement.
We numerically find that in the available parameter space and around the boundary, the axion always gets trapped by the QCD potential or the axion-dyon potential before the backreaction from PR becomes significant, even if the initial value of $\delta \theta_0$ is as large as 0.5.
The upper limit on $m_f/m_W$ given by the cyan shaded region comes from the relic density of dark fermion and depends on the choice of $m_W$, as discussed in the next section.
In addition, astrophysical bound provides a lower limit on $f_a \simeq 4 \times 10^{8}~\rm GeV$, assuming the KSVZ axion~\cite{Leinson:2021ety,Buschmann:2021juv}, as shown in the blue-shaded region.
One can alternatively modify the PQ charges of the SM particles to suppress the axion coupling to nucleons, electrons, and muons, thus lowering the bound to be around $10^7~\rm GeV$~\cite{DiLuzio:2017ogq,Bjorkeroth:2019jtx,Badziak:2023fsc}.

The acoustic misalignment mechanism also contributes to axion dark matter via nonzero modes. In the case where $\rho_\theta$ never dominates the energy density,%
\footnote{The condition for $\rho_\theta$ to never dominate is given by $Y_\theta^2 < \sqrt{1215/128 g_*} f_a/(\pi N_{\rm DW} m_S)$. We find some allowed parameter space in $m_S$ consistent with this assumption.}
the yield can be expressed as~\cite{Bodas:2025eca}
\begin{equation}
    Y_a \equiv \frac{n_a}{s} = \left. \frac{P_\zeta \, \dot\theta^2 f_a^2}
{4 H  s} \right|_{T = T_{S}} \, ,
\end{equation}
where $T_{S}$ refers to the temperature when
the radius of the rotation reaches the minimum around $f_a$,
\begin{equation}
T_{S} = \left(\frac{45}{2 \pi^2 g_*} \frac{N_{\rm DW} m_S f_a^2}{Y_\theta}\right)^{\frac{1}{3}}  \simeq 2 ~{\rm TeV} \left(\frac{c_B}{0.3}\right)^{\frac{1}{3}}  \left(\frac{N_{\rm DW} m_S}{10 ~{\rm MeV}}\right)^{\frac{1}{3}}, 
\end{equation}
with $N_{\rm DW}$ the domain wall number and $m_S$ the radial mode mass. In the second equality, we have used Eq.~(\ref{eq:YB from Ytheta}) to obtain $Y_\theta$.
The angular velocity at $T_{S}$ is $\dot\theta (T_{S}) = N_{\rm DW} m_S$ as determined by the equation of motion. In the absence of dissipation, this results in a contribution to the axion abundance of 
\begin{equation}
\label{eq:rhoAMM}
\left. \frac{\rho_a}{s} \right|_{\rm AMM} \simeq 0.4 m_a \left(\frac{\sqrt{g_*} N_{\rm DW} m_S Y_\theta^5}{f_a^4}  \right)^{\frac{1}{3}} M_{\rm Pl} P_\zeta .
\end{equation}

The amplitude of the fluctuation $\delta \theta$ at $T_{S}$ is given by $P_\zeta N_{\rm DW}^2 m_S^2 f_a^2 = \delta \theta_{S}^2 k^2 f_a^2$, with the dominant mode $k=H(T_{S})$, and redshifts as $\delta \theta \propto T$. At the time of zero-mode dissipation $T_{\rm di}$, we would have $\delta\theta_{\rm di} = \delta \theta_{S} T_{\rm di} / T_{S}$ without nonzero-mode dissipation. If $\delta\theta_{\rm di} > O(1)$, dissipation would actually reduce $\delta\theta_{\rm di}$ to $O(1)$ according to Sec.~\ref{subsec:nonzero_di}, suppressing the AMM contribution in Eq.~(\ref{eq:rhoAMM}) by a factor of $1/\delta\theta_{\rm di}^2$. The final contribution with nonzero-mode dissipation is then given by $\left. \rho_a/s \right|_{\rm AMM} \times \min (1,1/\delta\theta_{\rm di}^2)$.

For the parameter space we consider, we find that, with $m_S \simeq 10$ MeV, the undissipated abundance $\left. \rho_a/s \right|_{\rm AMM}$ overproduces DM by a factor of $O(100)$, while the undissipated $\delta\theta_{\rm di} = O(100)$. With $\delta\theta_{\rm di}$ dissipated down to $O(1)$, the AMM contribution to dark matter becomes negligible.%
\footnote{
With $m_S\sim 10$ MeV, in the KSVZ model~\cite{Kim:1979if,Shifman:1979if}, $S$ decouples from the thermal bath when it is relativistic and decays into axions much after becoming non-relativistic.
Too much dark radiation is produced. By coupling the PQ breaking field with the SM Higgs, $S$ may be kept in equilibrium with the thermal bath when it becomes non-relativistic, so that the constraint from dark radiation can be avoided. The required coupling with the Higgs can be probed by NA62 and KLEVER~\cite{Harigaya:2019qnl,NA62:2017rwk,KLEVERProject:2019aks}.
}

\subsection{Dark Fermion Dark Matter}
Besides the axion and monopole, the dark fermion may also contribute to the DM relic density.
Since the monopole energy density is large, the number density of the created dark fermions is much larger than their equilibrium value.
These fermions then annihilate into massless dark photons with the following $s$-wave cross section at the tree-level,
\begin{align}
     (\sigma v)_0 \simeq \frac{\pi \alpha_D^2}{16m_f^2}\,.
\end{align}
Once the annihilation process freezes out, the dark fermion relic contributes to DM.
This annihilation cross section is altered by the Sommerfeld effect~\cite{Sommerfeld:1931qaf} due to the non-perturbative long-range potential mediated by the dark photon.
The Sommerfeld factor for a massless mediator is~\cite{Arkani-Hamed:2008hhe,Cassel:2009wt}
\begin{align}
    S(v) = \frac{\pi \alpha_D / (2v) }{1 - \exp(-\pi \alpha_D/(2v))}\,,
\end{align}
where $v$ is the relative velocity between initial state particles in the center of mass frame.
Since the self-interaction among fermions is still active, kinetic equilibrium is maintained.
The thermal averaged cross section is then calculated as
\begin{align}
    \langle \sigma v \rangle = \left( \frac{x}{4 \pi} \right)^{3/2} \int_0^\infty 4 \pi v^2 \exp\left(\frac{- x v^2}{4}\right) (\sigma v)_0 S(v) d v\,,
\end{align}
where $x\equiv m_f/T$.

To build up an intuition, we first present the tree-level result here with an analytical approach.
Freeze-out happens when $n_f \langle \sigma v \rangle = 3H$.
The relic abundance is thus set by
\begin{equation}
    n_f \simeq \frac{\Gamma_\theta}{\langle \sigma v \rangle} =  \frac{16}{\pi^2 \alpha_D} f_M \xi_{\rm DM} \left( \frac{c_B}{Y_B} \right) \left( \frac{T_{\rm EW}}{f_a} \right)^2  m_W^2 \left( \frac{m_f}{m_W} \right)^3\,.
\end{equation}
Here we use the fact that $\Gamma_\theta = 3 H$ at $T=T_{\rm di}$.
The relic energy density compared to the DM density is solved as
\begin{equation}
\label{eq:fermi DM}
    \frac{\Omega_f}{\Omega_{\rm DM}} \sim  ~ 0.17 \left(\frac{f_a}{ 10^9~\mathrm{GeV}}\right) \left(\frac{m_f}{0.01 m_W}\right)^{5/2}
     \left(\frac{m_W}{20~\mathrm{TeV}}\right)^{3} \left(\frac{0.3}{c_B}\right)^{1/2} \left(\frac{0.2}{\alpha_D}\right)^{5/2} \left(\frac{0.5}{f_M}\right)^{1/2}\,.
\end{equation}
Although the Sommerfeld effect can change the quantitative result, this equation shows that $m_f$ is bounded from above, and the constraint becomes stronger for larger $m_W$. 

To obtain an accurate result, we numerically solve the following Boltzmann equation governing the dark fermion abundance. There are two $U(1)_D$ charged Dirac fermions. Each fermion follows
\begin{align}
    \frac{dY}{dx} &= \frac{- \langle \sigma v \rangle/2}{H x} s (Y^2 - Y_{\rm eq}^2) \\
    Y_{\rm eq} &= \frac{45}{2\pi^4} \left(\frac{\pi}{8}\right)^{1/2} \frac{g_f}{g_*} x^{3/2}e^{-x}\,,
\end{align}
where $Y \equiv n_f/s$ is the yield of the dark fermion, $g_f = 4$ is the degree of freedom of the dark fermion.
The factor of $1/2$ on the cross section comes from the fact that the dark fermion is Dirac.
We make the approximation that the creation of fermions from level crossing happens at the moment of $t_{\rm di}$.
The fermion number density at this moment is 
\begin{align}
    n_f (T_{\rm di})  = n_M \times \frac{\dot{\theta} t_{\rm di}}{2\pi}\,,
\end{align}
where the last factor represents the number of periods that the axion rotation has passed, with level crossing occurring once per period.
The initial condition of the Boltzmann equation is set as $Y(m_f/T_{\rm di}) = n_f (T_{\rm di}) /s(T_{\rm di})$.
We then multiply the resultant density by a factor of 2 to take into account the presence of two Dirac fermions.

Fig.~\ref{fig:main} shows the DM overproduction constraint in the green shaded region.
The fermion DM contribution imposes an upper bound on $m_f/m_W$, and is the only constraint that depends on the scale of $m_W$.
As illustrated in Fig.~\ref{fig:main}, increasing $m_W$ from 20 TeV to 100~TeV raises the fermion DM abundance and therefore lowers the allowed value of $m_f/m_W$, shrinking the viable parameter space.

At last, we comment on the parameter dependence of $f_M$.
Reducing $f_M$ allows more energy density from axion and fermion DM, which tends to open the parameter space.
However, a smaller $f_M$ also increases fermion and axion DM, as demonstrated in Eqs.~\eqref{eq:axion DM} and~\eqref{eq:fermi DM}.
The former effect is linear, while the latter becomes dominant when $f_M$ is small.
As a result, $f_M$ is optimized around intermediate values.
Numerically, we find the optimal value to be $f_M \simeq 0.5$.

\section{Conclusion}
\label{sec:conclusion}
We develop a novel mechanism in which a rotating QCD axion dissipates its kinetic energy via its interactions with ’t Hooft–Polyakov monopoles from a dark sector. This allows the QCD axion to generate the baryon asymmetry via axiogenesis without overproducing axion dark matter via kinetic or acoustic misalignment, while dark monopoles also constitute a fraction of dark matter. The axion’s coupling to the dark $U(1)$ gauge field turns monopoles into dyons via the Witten effect, and their electric charge evolves as $\theta$ increases. When the energy gaps meet the dark fermion mass, repeated crossings on quantized dyonic levels emit dark fermions and dissipate the axion kinetic energy. This mechanism preserves the baryon asymmetry generated by axiogenesis while solving the axion overproduction problem from kinetic misalignment in the minimal scenario.

In our setup, dark matter is composed of dark magnetic monopoles, dark fermions produced from these level crossing events, and axions produced by kinetic misalignment. In the viable parameter space, these three components have similar energy densities.

Our mechanism exhibits several predictions that may be testable in the future. It prefers a dark fermion mass of several hundred GeV to successfully dissipate the axion kinetic energy and not overproduce dark fermions. Also, the decay constant of the QCD axion is required to be below $10^9$ GeV, providing a specific target for axion searches. The dark magnetic monopoles and dark fermions will be self-interacting due to the dark $U(1)$ force, which comes with astrophysical signals for dissipative self-interacting dark matter~\cite{Agrawal:2016quu,Chang:2018bgx,Essig:2018pzq,Huo:2019yhk,Shen:2021frv,Xiao:2021ftk,Shen:2022opd}.

\section*{Acknowledgement} We are grateful to Lisa Randall for useful discussions.
The discussion of this project was initiated during PIKIMO Fall 2024 at the University of Michigan, Ann Arbor.
Fermilab is operated by Fermi Forward Discovery Group, LLC under Contract No. 89243024CSC000002 with the U.S. Department of Energy, Office of Science, Office of High Energy Physics.
This work was supported by the U.S. Department of Energy under Grant No.~DE-SC0025611 (RC), No.~DE-SC0009924 (KH), and No.~DE-SC0026297 (HX); DOE distinguished scientist fellowship grant FNAL 22-33 (IRW); and the World Premier International Research Center Initiative (WPI), MEXT, Japan (Kavli IPMU) (KH). For facilitating portions of this research, RC wishes to acknowledge the Center for Theoretical Underground Physics and Related Areas (CETUP*), the Institute for Underground Science at Sanford Underground Research Facility (SURF), and the South Dakota Science and Technology Authority for hospitality and financial support, as well as for providing a stimulating environment when part of this work was performed.

\bibliographystyle{JHEP}
\bibliography{main}

\end{document}